\documentclass[twocolumn,showpacs,preprintnumbers,superscriptaddress,amsmath,amssymb,pre]{revtex4-1}

\usepackage{graphicx}
\usepackage{dcolumn}
\usepackage{bm}
\usepackage{color}
\usepackage{multirow}
\usepackage{listings}

\begin{document}

\preprint{PRE/MFDMA}

\title{Detrending moving average algorithm for multifractals}

\author{Gao-Feng Gu}
 \affiliation{School of Business, East China University of Science and Technology, Shanghai 200237, China} %
 \affiliation{Research Center for Econophysics, East China University of Science and Technology, Shanghai 200237, China} %

\author{Wei-Xing Zhou}
 \email{wxzhou@ecust.edu.cn}
 \affiliation{School of Business, East China University of Science and Technology, Shanghai 200237, China} %
 \affiliation{Research Center for Econophysics, East China University of Science and Technology, Shanghai 200237, China} %
 \affiliation{School of Science, East China University of Science and Technology, Shanghai 200237, China} %
 \affiliation{Engineering Research Center of Process Systems Engineering (Ministry of Education), East China University of Science and Technology, Shanghai 200237, China} %
 \affiliation{Research Center on Fictitious Economics \& Data Science, Chinese Academy of Sciences, Beijing 100080, China} %

\date{\today}

\begin{abstract}
The detrending moving average (DMA) algorithm is a widely used technique to quantify the long-term correlations of non-stationary time series and the long-range correlations of fractal surfaces, which contains a parameter $\theta$ determining the position of the detrending window. We develop multifractal detrending moving average (MFDMA) algorithms for the analysis of one-dimensional multifractal measures and higher-dimensional multifractals, which is a generalization of the DMA method. The performance of the one-dimensional and two-dimensional MFDMA methods is investigated using synthetic multifractal measures with analytical solutions for backward ($\theta=0$), centered ($\theta=0.5$), and forward ($\theta=1$) detrending windows. We find that the estimated multifractal scaling exponent $\tau(q)$ and the singularity spectrum $f(\alpha)$ are in good agreement with the theoretical values. In addition, the backward MFDMA method has the best performance, which provides the most accurate estimates of the scaling exponents with lowest error bars, while the centered MFDMA method has the worse performance. It is found that the backward MFDMA algorithm also outperforms the multifractal detrended fluctuation analysis (MFDFA). The one-dimensional backward MFDMA method is applied to analyzing the time series of Shanghai Stock Exchange Composite Index and its multifractal nature is confirmed.
\end{abstract}


\maketitle

\section{Introduction}
\label{sec:Introduction}

Fractals and multifractals are ubiquitous in natural and social sciences \cite{Mandelbrot-1983,Mandelbrot-1997,Sornette-2004}. There are a large number of methods developed to characterize the properties of fractals and multifractals. The classic method is the Hurst analysis or rescaled range analysis (R/S) for time series \cite{Hurst-1951-TASCE,Mandelbrot-Wallis-1969b-WRR} and fractal surfaces \cite{AlvarezRamirez-Echeverria-Rodriguez-2008-PA}. The wavelet transform module maxima (WTMM) method is a more powerful tool to address the multifractality
\cite{Holschneider-1988-JSP,Muzy-Bacry-Arneodo-1991-PRL,Muzy-Bacry-Arneodo-1993-JSP,Muzy-Bacry-Arneodo-1993-PRE,Muzy-Bacry-Arneodo-1994-IJBC}, even for high-dimensional multifractal measures in the fields of image technology and three-dimensional turbulence
\cite{Arneodo-Decoster-Roux-2000-EPJB,Decoster-Roux-Arneodo-2000-EPJB,Roux-Arneodo-Decoster-2000-EPJB,Kestener-Arneodo-2003-PRL,Kestener-Arneodo-2004-PRL}.
Another popular method is the detrended fluctuation analysis (DFA), which has the advantages of easy implementation and robust estimation even for short signals
\cite{Taqqu-Teverovsky-Willinger-1995-Fractals,Montanari-Taqqu-Teverovsky-1999-MCM,Audit-Bacry-Muzy-Arneodo-2002-IEEEtit}.
The DFA method was originally invented to study the long-range dependence in coding and noncoding DNA nucleotides sequence
\cite{Peng-Buldyrev-Havlin-Simons-Stanley-Goldberger-1994-PRE} and then applied to time series in various fields \cite{Gu-Zhou-2007-PA,Gu-Chen-Zhou-2007-EPJB,Gu-Chen-Zhou-2008c-PA,Jiang-Chen-Zhou-2009-PA}.
The DFA algorithm was extended to analyze the multifractal time series, which is termed as multifractal detrended fluctuation analysis (MFDFA)
\cite{Kantelhardt-Zschiegner-KoscielnyBunde-Havlin-Bunde-Stanley-2002-PA}. These DFA and MFDFA methods were also generalized to analyze high-dimensional fractals and multifractals \cite{Gu-Zhou-2006-PRE}.

A more recent method is based on the moving average (MA) or mobile average technique \cite{Carbone-2009-IEEE}, which was first proposed by Vandewalle and Ausloos to estimate the Hurst exponent of self-affinity signals \cite{Vandewalle-Ausloos-1998-PRE} and further developed to the detrending moving average (DMA) by considering the second-order difference between the original signal and its moving average function \cite{Alessio-Carbone-Castelli-Frappietro-2002-EPJB}. Because the DMA method can be easily implemented to estimate the correlation properties of non-stationary series without any assumption, it is widely applied to the analysis of real-world time series \cite{Carbone-Castelli-2003-SPIE,Carbone-Castelli-Stanley-2004-PA,Carbone-Castelli-Stanley-2004-PRE,Varotsos-Sarlis-Tanaka-Skordas-2005-PRE,Serletis-Rosenberg-2007-PA,Arianos-Carbone-2007-PA,Matsushita-Gleria-Figueiredo-Silva-2007-PLA,Serletis-Rosenberg-2009-CSF} and synthetic signals \cite{Carbone-Stanley-2004-PA,Xu-Ivanov-Hu-Chen-Carbone-Stanley-2005-PRE,Serletis-2008-CSF}. Recently, Carbone extended the one-dimensional DMA method to higher dimensions to estimate the Hurst exponents of higher-dimensional fractals \cite{Carbone-2007-PRE,Turk-Carbone-Chiaia-2010-PRE}. Extensive numerical experiments unveil that the performance of the DMA method are comparable to the DFA method with slightly different priorities under different situations \cite{Xu-Ivanov-Hu-Chen-Carbone-Stanley-2005-PRE,Bashan-Bartsch-Kantelhardt-Havlin-2008-PA}.

In this paper, we extend the DMA method to multifractal detrending moving average (MFDMA), which is designed to analyze multifractal time series and multifractal surfaces. Further extensions to higher-dimensional versions are straightforward. The performance of the MFDMA algorithms is investigated using synthetic multifractal measures with known multifractal properties. We also compare the performance of MFDMA with MFDFA, and find that MFDMA is superior to MFDFA for multifractal analysis.

The paper is organized as follows. In Sec.~\ref{S1:MFDMA:1D}, we describe the algorithm of one-dimensional MFDMA and show the results of numerical simulations. We also apply the one-dimensional MFDMA to analyze the time series of intraday Shanghai Stock Exchange Composite Index (SSEC). In Sec.~\ref{S1:MFDMA:2D}, we describe the algorithm of two-dimensional MFDMA and report the results of numerical simulations as well. We discuss and conclude in Sec.~\ref{S1:Summary}.

\section{One-dimensional multifractal detrending moving average analysis}
\label{S1:MFDMA:1D}

\subsection{Algorithm}
\label{S2:MFDMA:1D:Algo}

{\em{Step 1}}. Consider a time series $x(t)$, $t=1,2,\cdots,N$. We construct the sequence of cumulative sums
\begin{equation}
y(t)=\sum_{i=1}^{t}{x(i)}, ~~t=1, 2, \cdots, N.
 \label{Eq:1ddma_y}
\end{equation}

{\em{Step 2}}. Calculate the moving average function $\widetilde{y}(t)$ in a moving window \cite{Arianos-Carbone-2007-PA},
\begin{equation}
\widetilde{y}(t)=\frac{1}{n}\sum_{k=-\lfloor(n-1)\theta\rfloor}^{\lceil(n-1)(1-\theta)\rceil}y(t-k),
\label{Eq:1ddma_y1}
\end{equation}
where $n$ is the window size, $\lfloor{x}\rfloor$ is the largest integer not greater than $x$, $\lceil{x}\rceil$ is the smallest integer not smaller than $x$, and $\theta$ is the position parameter with the value varying in the range $[0,1]$. Hence, the moving average function considers $\lceil(n-1)(1-\theta)\rceil$ data points in the past and $\lfloor(n-1)\theta\rfloor$ points in the future. We consider three special cases in this paper. The first case $\theta=0$ refers to the backward moving average \cite{Xu-Ivanov-Hu-Chen-Carbone-Stanley-2005-PRE}, in which the moving average function $\widetilde{y}(t)$ is calculated over all the past $n-1$ data points of the signal. The second case $\theta=0.5$ corresponds to the centered moving average \cite{Xu-Ivanov-Hu-Chen-Carbone-Stanley-2005-PRE}, where $\widetilde{y}(t)$ contains half past and half future information in each window. The third case $\theta=1$ is called the forward moving average, where $\widetilde{y}(t)$ considers the trend of $n-1$ data points in the future.

{\em{Step 3}}. Detrend the signal series by removing the moving average function $\widetilde{y}(i)$ from $y(i)$, and obtain the residual sequence $\epsilon(i)$ through
\begin{equation}
\epsilon(i)=y(i)-\widetilde{y}(i),
\label{Eq:1ddma_epsilon}
\end{equation}
where $n-\lfloor(n-1)\theta\rfloor\leqslant{i}\leqslant{N-\lfloor(n-1)\theta\rfloor}$.

{\em{Step 4}}. The residual series $\epsilon(i)$ is divided into $N_n$ disjoint segments with the same size $n$, where $N_n=\lfloor{N}/n-1\rfloor$. Each segment can be denoted by $\epsilon_v$ such that $\epsilon_v(i)=\epsilon(l+i)$ for $1\leqslant{i}\leqslant{n}$, where $l=(v-1)n$. The root-mean-square function $F_v(n)$ with the segment size $n$ can be calculated by
\begin{equation}
F_v^2(n)=\frac{1}{n}\sum_{i=1}^{n}\epsilon_v^2(i).
\label{Eq:dma_m_F}
\end{equation}

{\em{Step 5}}. The $q$th order overall fluctuation function $F_q(n)$ is determined as follows,
\begin{equation}
  F_q(n) = \left\{\frac{1}{N_n}\sum_{v=1}^{N_n} {F_v^q(n)}\right\}^{\frac{1}{q}},
  \label{Eq:1ddma_Fqs}
\end{equation}
where $q$ can take any real value except for $q=0$. When $q=0$,
we have
\begin{equation}
  \ln[F_0(n)] = \frac{1}{N_n}\sum_{v=1}^{N_n}{\ln[F_v(n)]},
  \label{Eq:1ddma_Fq0}
\end{equation}
according to L'H\^{o}spital's rule.

{\em{Step 6}}. Varying the values of segment size $n$, we can determine the power-law relation between the function $F_q(n)$ and the size scale $n$, which reads
\begin{equation}
  F_q(n)\sim{n}^{h(q)}.
  \label{Eq:dma_m_h}
\end{equation}

According to the standard multifractal formalism, the multifractal scaling exponent $\tau(q)$ can be used to characterize the multifractal nature, which reads
\begin{equation}
\tau(q)=qh(q)-D_f,
\label{Eq:tau:hq}
\end{equation}
where $D_f$ is the fractal dimension of the geometric support of the multifractal measure \cite{Kantelhardt-Zschiegner-KoscielnyBunde-Havlin-Bunde-Stanley-2002-PA}. For time series analysis, we have $D_f=1$. If the scaling exponent function $\tau(q)$ is a nonlinear function of $q$, the signal has multifractal nature. It is easy to obtain the singularity strength function $\alpha(q)$ and the multifractal spectrum $f(\alpha)$ via the Legendre transform \cite{Halsey-Jensen-Kadanoff-Procaccia-Shraiman-1986-PRA}
\begin{equation}
    \left\{
    \begin{array}{ll}
        \alpha(q)={\rm{d}}\tau(q)/{\rm{d}}q\\
        f(q)=q{\alpha}-{\tau}(q)
    \end{array}
    \right..
\label{Eq:f:alpha:tau}
\end{equation}

\subsection{Numerical experiments}
\label{S2:MFDMA:1D:Numerical}

In the numerical experiments, we generate one-dimensional multifractal measure to investigate the performance of MFDMA, which is compared with MFDFA. We apply the $p$-model, a multiplicative cascading process, to synthesize the multifractal measure \cite{Meneveau-Sreenivasan-1987-PRL}. Starting from a measure $\mu$ uniformly distributed on an interval $[0,1]$. In the first step, the measure is redistributed on the interval, $\mu_{1,1}=\mu p_1$ to the first half and $\mu_{1,2}=\mu p_2=\mu(1-p_1)$ to the second half. One partitions it into two sub-lines with the same length. In the $(k+1)$-th step, the measure $\mu_{k,i}$ on each of the $2^k$ line segments is redistributed into two parts, where $\mu_{k+1,2i-1}=\mu_{k,i}p_1$ and  $\mu_{k+1,2i}=\mu_{k,i}p_2$. We repeat the procedure for $14$ times and finally generate the one-dimensional multifractal measure with the length $2^{14}$. In this paper, we present the results when the parameters are $p_1=0.3$ and $p_2=0.7$. The results for other parameters are qualitatively the same.

\begin{figure*}[htb]
  \centering
  \includegraphics[width=8cm]{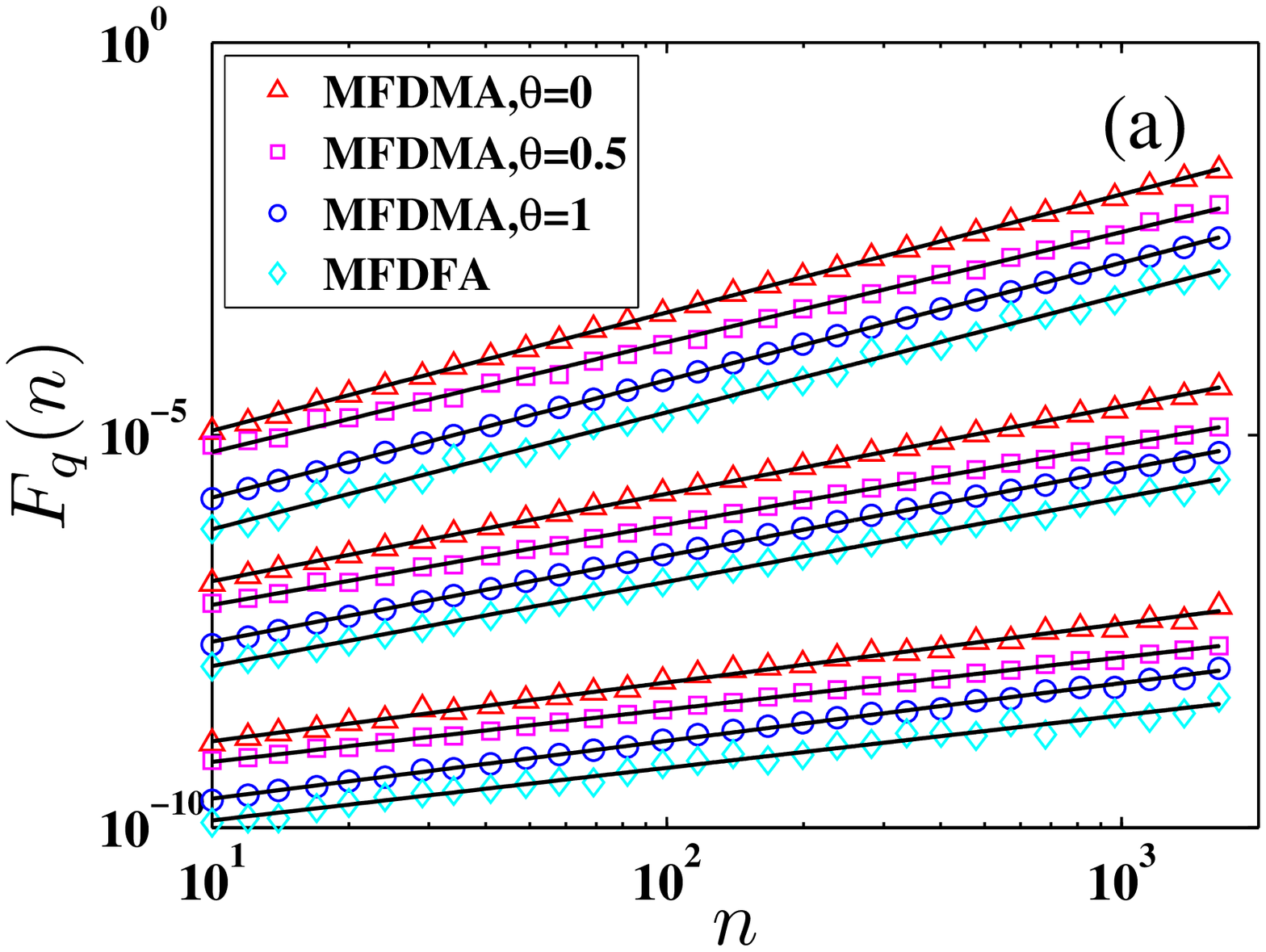}
  \includegraphics[width=8cm]{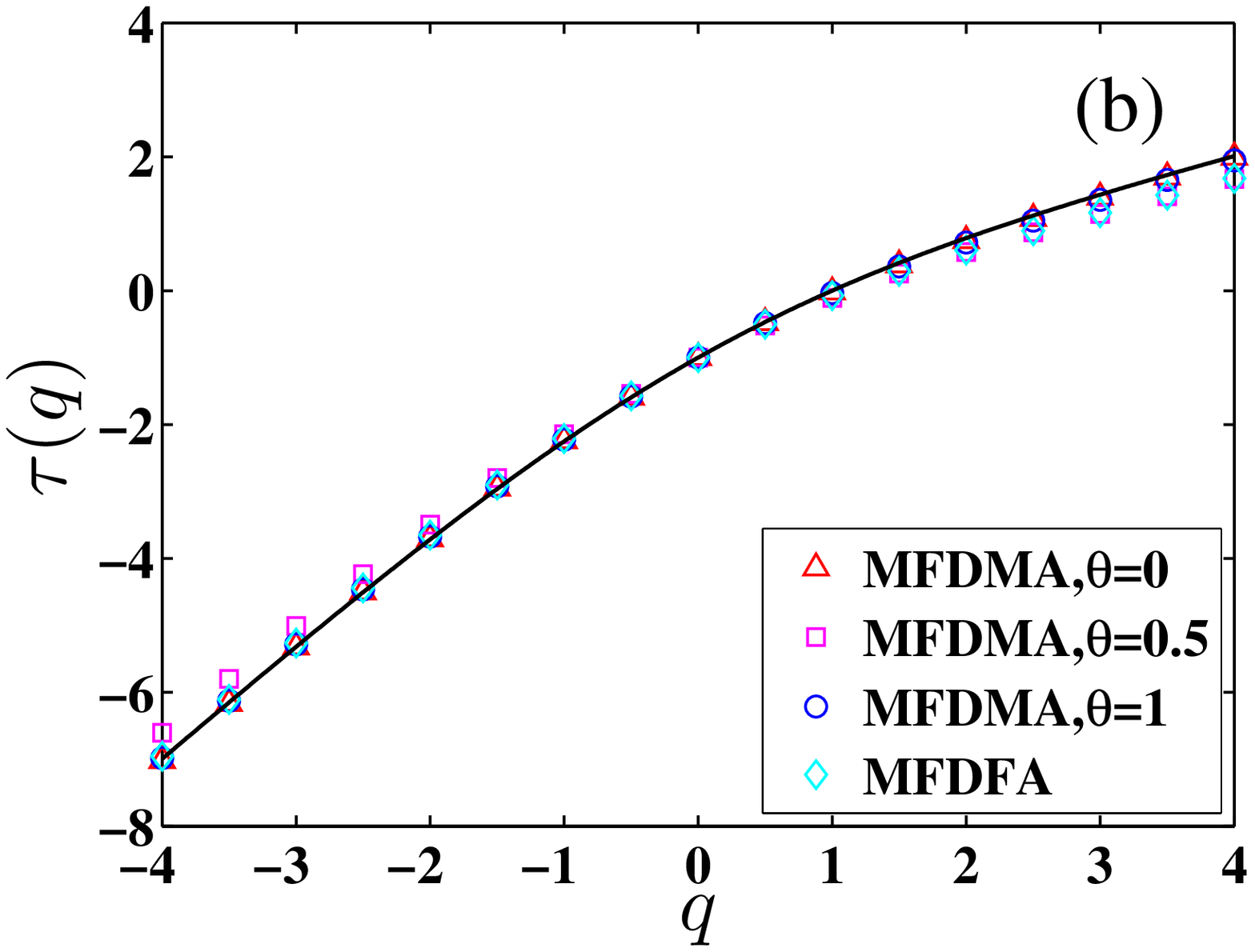}
  \includegraphics[width=8cm]{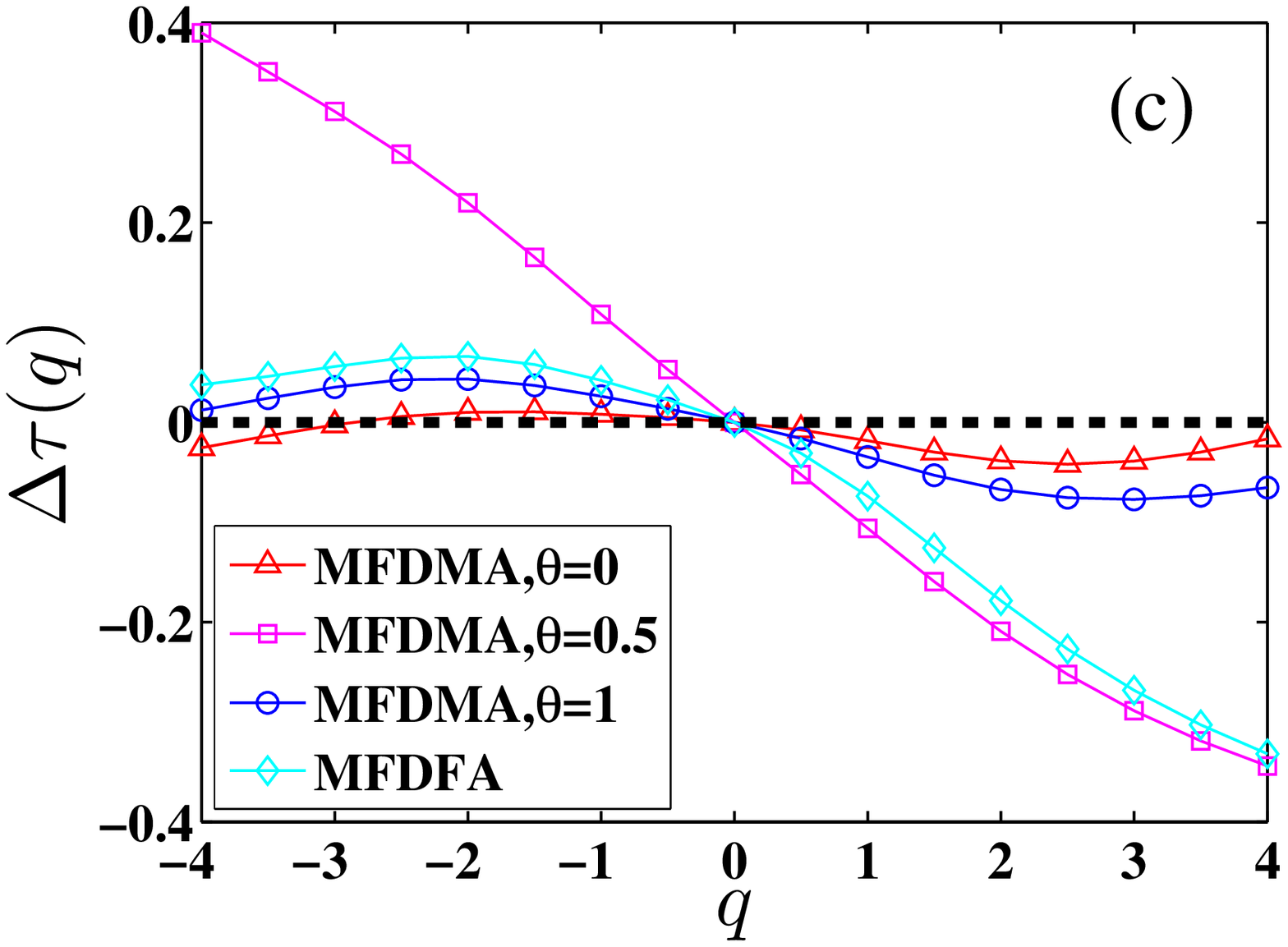}
  \includegraphics[width=8cm]{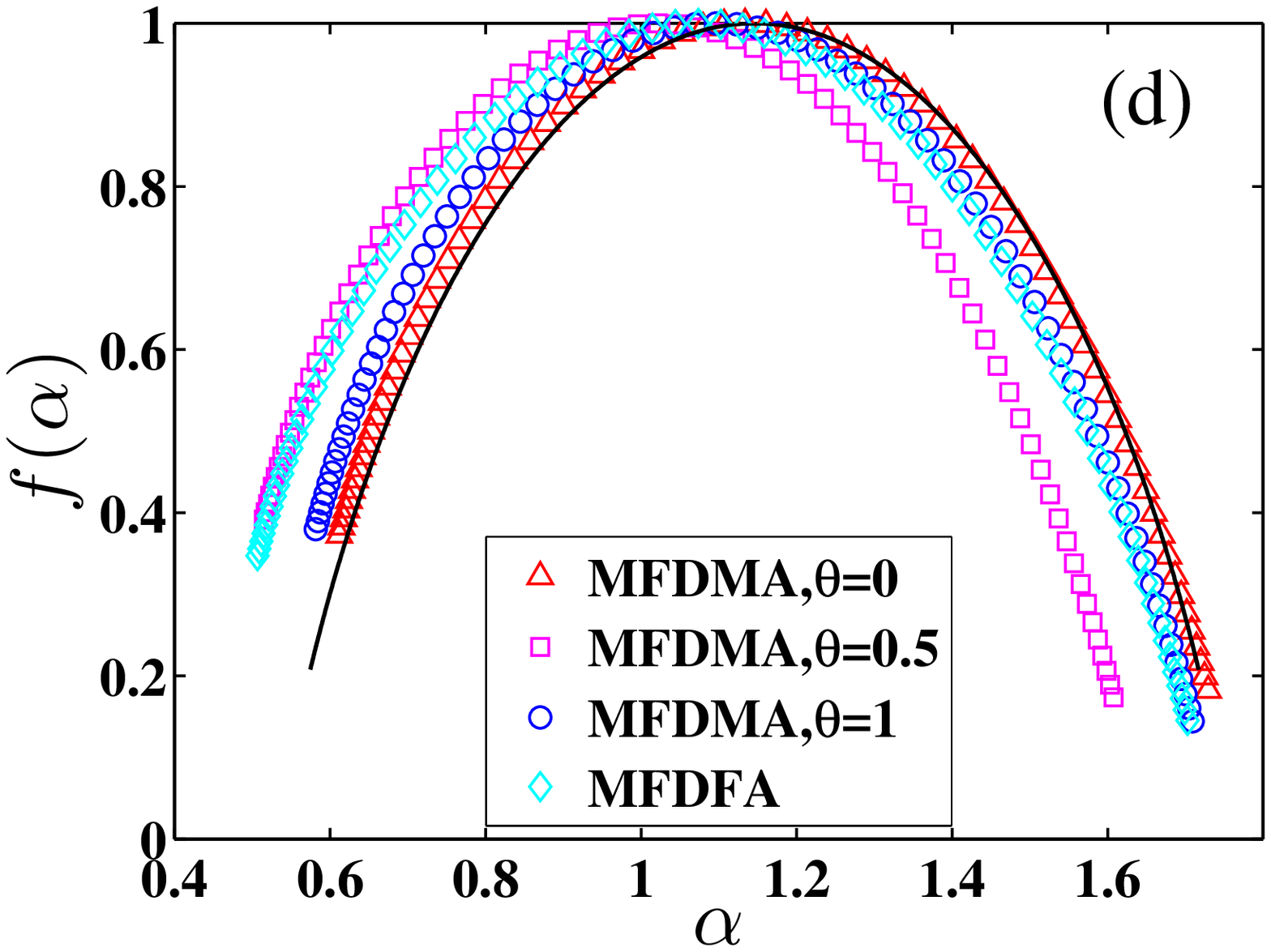}
  \caption{(Color online) Multifractal analysis of the one-dimensional multifractal binomial measure using the three MFDMA algorithms and the MFDFA approach. (a) Power-law dependence of the fluctuation functions $F_q(n)$ with respect to the scale $n$ for $q=-4$, $q=0$, and $q=4$. The straight lines are the best power-law fits to the data. The results have been translated vertically for better visibility. (b) Multifractal mass exponents $\tau(q)$ obtained from the MFDMA and MFDFA methods with the theoretical curve shown as a solid line. (c) Differences $\Delta{\tau}(q)$ between the estimated mass exponents and their theoretical values for the four algorithms. (d) Multifractal spectra $f(\alpha)$ with respect to the singularity strength $\alpha$ for the four methods. The continuous curve is the theoretical multifractal spectrum.}
  \label{Fig:1DMFDMA}
\end{figure*}

\begin{figure*}[htb]
  \centering
  \includegraphics[width=8.3cm]{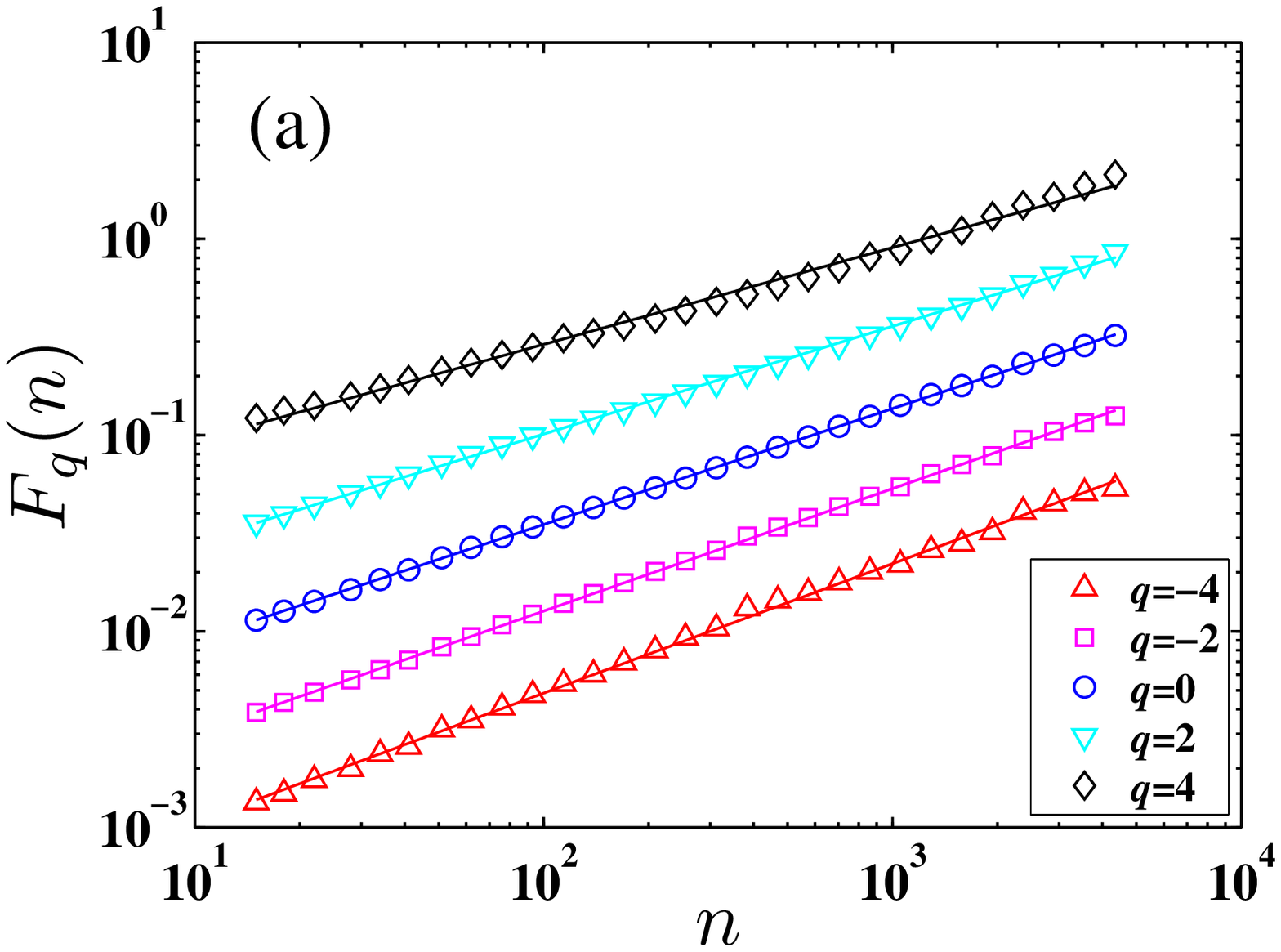}
  \includegraphics[width=8cm]{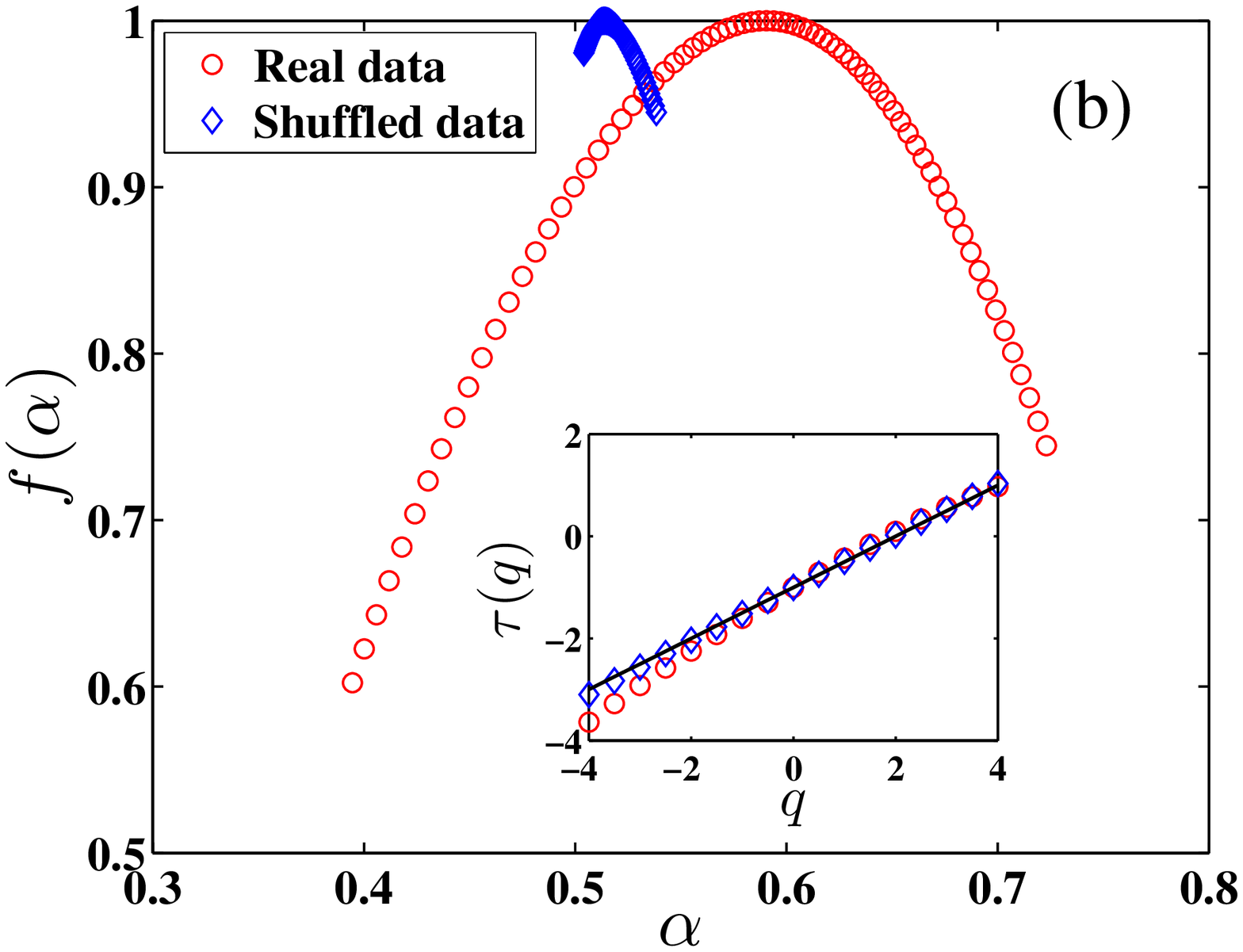}
  \caption{(Color online) Multifractal analysis of the 5-min return time series of the SSEC index using the backward MFDMA method. (a) Power-law dependence of the fluctuation functions $F(n)$ with respect to the scale $n$. The solid lines are the least-squares fits to the data. The results corresponding to $q=-2$, $q=0$, $q=2$ and $q=4$ have been translated vertically for clarity. (b) Multifractal spectra $f(\alpha)$ of the raw return series of SSEC and its shuffled series. Inset: Multifractal scaling exponents $\tau(q)$ as a function of $q$.}
  \label{Fig:1DMFDMA:SSEC}
\end{figure*}

We calculate the fluctuation function $F_q(n)$ of the synthetical multifractal measure using the MFDMA and MFDFA methods, and present the fluctuation function $F_q(n)$ in Fig.~\ref{Fig:1DMFDMA}(a). We find that the function $F_q(n)$ well scales with the scale size $n$. Using the least squares fitting method, we obtain the slopes $h(q)$ for MFDMA ($\theta=0$, $\theta=0.5$ and $\theta=1$) and MFDFA respectively, which are illustrated in Table~\ref{Tb:MFDMA:1D}. It is found that the error bars of the three MFDMA algorithms are all smaller than the MFDFA method, which implies that it is easier to determine the scaling ranges for the MFDMA algorithms. In most cases, the algorithms underestimate the $h(q)$ values and the backward MFDMA approach gives the best estimates. There is an interesting feature in Fig.~\ref{Fig:1DMFDMA}(a) showing evident log-periodic oscillations in the MFDFA $F_q(n)$ curves, which is intrinsic for the multifractal binomial measure \cite{Zhou-Sornette-2009b-PA}.

\begin{table}[htp]
 \centering
 \caption{The MFDMA exponents $h(q)$ for $q=-4$, -2, 0, 2, and 4 of the one-dimensional synthetic multifractal measure with the parameters $p_1=0.3$ and $p_2=0.7$ using the MFDMA ($\theta=0$, $\theta=0.5$ and $\theta=1$) and MFDFA methods. The numbers in the parentheses are the standard errors of the regression coefficient estimates using the $t$-test at the 5\% significance level.}
 \medskip
 \label{Tb:MFDMA:1D}
 \centering
 \begin{tabular}{cccccccccccc}
 \hline\hline
  \multirow{3}*[2mm]{$q$}&& \multicolumn{3}{c}{MFDMA}&& \multirow{3}*[2mm]{MFDFA} & \multirow{3}*[2mm]{Analytic} \\  %
  \cline{3-5}
     && $\theta=0$ & $\theta=0.5$ & $\theta=1$  \\
 \hline
    -4 && 1.505(4) & 1.401(12) & 1.496(2) && 1.490(17) & 1.499\\
    -2 && 1.354(3) & 1.249( 8) & 1.337(4) && 1.326( 9) & 1.359\\
     0 && 1.114(4) & 1.022( 5) & 1.096(5) && 1.074( 6) & 1.126\\
     2 && 0.874(6) & 0.788( 3) & 0.859(5) && 0.804(11) & 0.893\\
     4 && 0.749(9) & 0.667( 4) & 0.736(6) && 0.670(15) & 0.753\\
 \hline\hline
 \end{tabular}
\end{table}

We plot the multifractal scaling exponents $\tau(q)$ obtained from MFDMA ($\theta=0$, $\theta=0.5$ and $\theta=1$) and MFDFA in Fig.~\ref{Fig:1DMFDMA}(b). The theoretical formula of $\tau(q)$ of the multifractal measure generated by the $p$-model discussed above can be expressed by \cite{Halsey-Jensen-Kadanoff-Procaccia-Shraiman-1986-PRA}
\begin{equation}
  \tau_{\rm{th}}(q)=-\frac{\ln(p_1^q+p_2^q)}{\ln2},
  \label{Eq:tau:hq_2}
\end{equation}
which has been illustrated in Fig.~\ref{Fig:1DMFDMA}(b) as well. In order to quantitatively evaluate the performance of MFDMA and MFDFA, we calculate the relative estimation errors of the numerical values of $\tau(q)$ in reference to the corresponding theoretical values $\tau_{\rm{th}}(q)$
\begin{equation}
  \Delta{\tau}(q) = \tau(q)-\tau_{\rm{th}}(q),
  \label{Eq:MFDMA:dTau}
\end{equation}
which are shown in Fig.~\ref{Fig:1DMFDMA}(c). When $0<q\leqslant4$, all the four methods underestimate the $\tau(q)$ exponents. In contrast, when $-4\leqslant q <0$, these methods overestimate the scaling exponents with a few exceptions for the backward MFDMA case. It is evident that the backward MFDMA method ($\theta=0$) gives the most accurate estimation of the exponents, the forward MFDMA method ($\theta=1$) has the second best performance, and the centered MFDMA method ($\theta=0.5$) performs worst. We stress that both the backward and the forward MFDMA methods outperform the MFDFA approach.

According to the Legendre transform, we can numerically calculate the singularity strength functions $\alpha(q)$ and the multifractal spectrum functions $f(\alpha)$ with Eq.~(\ref{Eq:f:alpha:tau}) for the four methods, which are depicted in Fig.~\ref{Fig:1DMFDMA}(d). We also show the theoretical singularity spectrum as a continuous curve for comparison, where the singularity strength function $\alpha(q)$ can be calculated as follows
\begin{equation}
  \alpha_{\rm{th}}(q)=-\frac{p_1^q\ln{p_1}+p_2^q\ln{p_2}}{(p_1^q+p_2^q)\ln2},
  \label{Eq:1ddma:alpha}
\end{equation}
and the multifractal spectrum is
\begin{equation}
  f_{\rm{th}}(q) = -\frac{qp_1^q\ln{p_1}+qp_2^q\ln{p_2}}{(p_1^q+p_2^q)\ln2}+\frac{\ln(p_1^q+p_2^q)}{\ln2}.
  \label{Eq:1ddma:f}
\end{equation}
It also shows that the order of algorithm performance is the following: backward MFDMA, forward MFDMA, MFDFA, and centered MFDMA.

\subsection{Application to intraday SSEC time series}
\label{S2:MFDMA:1D:Appl}

We now apply the one-dimensional backward MFDMA method to the study of the multifractal properties of 5-min return series of Shanghai Stock Exchange Composite Index (SSEC) within the time period from January 2003 to April 2008. The number of data points is about $10^5$. In Fig.~\ref{Fig:1DMFDMA:SSEC}(a), we present the fluctuation function $F_q(n)$ with respect to the scale $n$. It is found that the function $F_q(n)$ and the scale $n$ have a sound power-law relation. The MFDMA exponents $h(q)$ can be estimated by the slopes of the straight lines illustrated in Fig.~\ref{Fig:1DMFDMA:SSEC}(a) for different $q$. Using the least-squares fitting method, we have $h(-4)=0.660\pm0.005$, $h(-2)=0.624\pm0.002$,  $h(0)=0.591\pm0.001$, $h(2)=0.531\pm0.003$, and $h(4)=0.493\pm0.008$. Since $h(2)$ is close to 0.5, it is confirmed that the return time series of the SSEC is almost uncorrelated, which is consistent with the earlier empirical results.

Figure \ref{Fig:1DMFDMA:SSEC}(b) shows the multifractal spectrum $f(\alpha)$ of return series of SSEC. The strength of multifractality can be characterized by the span of the multifractal singularity strength function, that is, $\Delta\alpha=\alpha_{\max}-\alpha_{\min}$. If $\Delta\alpha$ is large, the return series owns multifractality, while the return series is almost monofractal if $\Delta\alpha$ gets close to zero. We observe in the figure that $\Delta\alpha=0.72-0.39=0.33$, which means that the return series of SSEC has multifractal nature. The nonlinearity of the $\tau(q)$ curve in the inset confirms the presence of multifractality in the return series.

We shuffle the return time series and reconstruct a surrogate index. We then perform backward MFDMA analysis of the surrogate index. The results are also illustrated in Fig.~\ref{Fig:1DMFDMA:SSEC}(b). We find that the singularity spectrum shrinks much and the $\tau(q)$ function is almost linear, which implies that the fat-tailedness of the returns plays a minor role in producing the observed multifractality and the correlation structure is the main cause of multifractality. This observation is quite interesting since it is different from the conclusion that the fat-tailedness of the returns is the main origin of multifractality according to the MFDFA method \cite{Zhou-2009-EPL}.

\section{Two-dimensional multifractal detrending moving average analysis}
\label{S1:MFDMA:2D}

\subsection{Algorithm}
\label{S2:MFDMA:2D:Algo}

The two-dimensional MFDMA algorithm is used to investigate possible multifractal properties of surfaces, which can be denoted by a two-dimensional matrix $X(i_1,i_2)$ with $i_1=1,2,\cdots,N_1$ and $i_2=1,2,\cdots,N_2$. The algorithm is described as follows.

{\em{Step 1}}. Calculate the sum $Y(i_1,i_2)$ in a sliding window with size $n_1\times{n_2}$, where $n_1\leqslant{i_1}\leqslant{N_1-\lfloor{(n_1-1)\theta_1}\rfloor}$ and $n_2\leqslant{i_2}\leqslant{N_2-\lfloor{(n_2-1)\theta_2}\rfloor}$. The two position parameters $\theta_1$ and $\theta_2$ vary in the range $[0,1]$. Specifically, we extract a sub-matrix $Z(u_1,u_2)$ with size $n_1\times{n_2}$ from the matrix $X$, where $i_1-n_1+1\leqslant{u_1}\leqslant{i_1}$ and $i_2-n_2+1\leqslant{u_2}\leqslant{i_2}$. We can calculate the sum $Y(i_1,i_2)$ of $Z$ as follows,
\begin{equation}
  Y(i_1,i_2)=\sum_{j_1=1}^{n_1}\sum_{j_2=1}^{n_2}{Z(j_1,j_2)}.
  \label{Eq:2ddma_y_ii}
\end{equation}

{\em{Step 2}}. Determine the moving average function $\widetilde{Y}(i_1,i_2)$, where $n_1\leqslant{i_1}\leqslant{N_1-\lfloor{(n_1-1)\theta_1}\rfloor}$ and $n_2\leqslant{i_2}\leqslant{N_2-\lfloor{(n_2-1)\theta_2}\rfloor}$. We first extract a sub-matrix $W(k_1,k_2)$ with size $n_1\times{n_2}$ from the matrix $X$, where $k_1-\lceil{(n_1-1)(1-\theta_1)}\rceil\leqslant{k_1}\leqslant{k_1+\lfloor{(n_1-1)\theta_1}\rfloor}$ and $k_2-\lceil{(n_2-1)(1-\theta_2)}\rceil\leqslant{k_2}\leqslant{k_2+\lfloor{(n_2-1)\theta_2}\rfloor}$. Then we calculate the cumulative sum $\widetilde{W}(m_1,m_2)$ of $W$,
\begin{equation}
  \widetilde{W}(m_1,m_2)=\sum_{d_1=1}^{m_1}\sum_{d_2=1}^{m_2}{W(d_1,d_2)},
  \label{Eq:2ddma_z1}
\end{equation}
where $1\leqslant{m_1}\leqslant{n_1}$ and $1\leqslant{m_2}\leqslant{n_2}$. The moving average function $\widetilde{Y}(i_1,i_2)$ can be calculated as follows,
\begin{equation}
  \widetilde{Y}(i_1,i_2)=\frac{1}{n_1n_2}\sum_{m_1=1}^{n_1}\sum_{m_2=1}^{n_2}{\widetilde{W}(m_1,m_2)}.
  \label{Eq:2ddma_y1}
\end{equation}

{\em{Step 3}}. Detrend the matrix by removing the moving average function $\widetilde{Y}(i_1,i_2)$ from $Y(i_1,i_2)$, and obtain the residual matrix $\epsilon(i_1,i_2)$ as follows,
\begin{equation}
  \epsilon(i_1,i_2)=Y(i_1,i_2)-\widetilde{Y}(i_1,i_2),
  \label{Eq:2ddma_epsilon}
\end{equation}
where $n_1\leqslant{i_1}\leqslant{N_1-\lfloor{(n_1-1)\theta_1}\rfloor}$ and $n_2\leqslant{i_2}\leqslant{N_2-\lfloor{(n_2-1)\theta_2}\rfloor}$.

{\em{Step 4}}. The residual matrix $\epsilon(i_1,i_2)$ is partitioned into $N_{n_1}\times{N_{n_2}}$ disjoint rectangle segments of the same size $n_1\times{n_2}$, where $N_{n_1}=\lfloor{(N_1-n_1(1+\theta_1))/n_1}\rfloor$ and $N_{n_2}=\lfloor{(N_2-n_2(1+\theta_2))/n_2}\rfloor$. Each segment can be denoted by $\epsilon_{v_1,v_2}$ such that $\epsilon_{v_1,v_2}(i_1,i_2)=\epsilon(l_1+i_1,l_2+i_2)$ for $1\leqslant{i_1}\leqslant{n_1}$ and $1\leqslant{i_2}\leqslant{n_2}$, where $l_1=(v_1-1)n_1$ and $l_2=(v_2-1)n_2$.
The detrended fluctuation $F_{v_1,v_2}(n_1,n_2)$ of segment $\epsilon_{v_1,v_2}(i_1,i_2)$ can be calculated as follows,
\begin{equation}
  F_{v_1,v_2}^2(n_1,n_2)=\frac{1}{n_1n_2}\sum_{i_1=1}^{n_1}\sum_{i_2=1}^{n_2}\epsilon_{v_1,v_2}^2(i_1,i_2).
  \label{Eq:2ddma_F}
\end{equation}

{\em{Step 5}}. The $q$th order overall fluctuation function $F_q(n)$ is calculated as follows,
\begin{equation}
  F_q(n) = \left\{\frac{1}{N_{n_1}N_{n_2}}\sum_{v_1=1}^{N_{n_1}}\sum_{v_2=1}^{N_{n_2}}{F_{v_1,v_2}^q(n_1,n_2)}\right\}^{\frac{1}{q}},
  \label{Eq:2ddma_Fq}
\end{equation}
where $n^2=\frac{1}{2}(n_1^2+n_2^2)$ and $q$ can take any real values except for $q=0$. When $q=0$, we have
\begin{equation}
  \ln[F_0(n)] = \frac{1}{N_{n_1}N_{n_2}}\sum_{v_1=1}^{N_{n_1}}\sum_{v_2=1}^{N_{n_2}}{\ln[F_{v_1,v_2}(n_1,n_2)]},
  \label{Eq:2ddma_Fq0}
\end{equation}
according to L'H\^{o}spital's rule.

{\em{Step 6}}. Varying the segment sizes $n_1$ and $n_2$, we are able to determine the power-law relation between the fluctuation function ${F_q(n)}$ and the scale $n$,
\begin{equation}
  F_q(n)\sim{n}^{h(q)},
  \label{Eq:2ddma_h}
\end{equation}

In this paper, we particularly adopt $n=n_1=n_2$ and $\theta=\theta_1=\theta_2$ for the isotropic implementation of the tow-dimensional MFDMA algorithm. Applying Eqs.~(\ref{Eq:tau:hq}) and (\ref{Eq:f:alpha:tau}), we can obtain the multifractal scaling exponent $\tau(q)$, the singularity strength function $\alpha(q)$ and the multifractal spectrum $f(\alpha)$, respectively. For the two-dimensional multifractal measures, we have $D_f=2$ in Eq.~(\ref{Eq:tau:hq}).

\subsection{Numerical experiments}
\label{S2:MFDMA:2D:Numerical}

\begin{figure*}[htb]
  \centering
  \includegraphics[width=8cm]{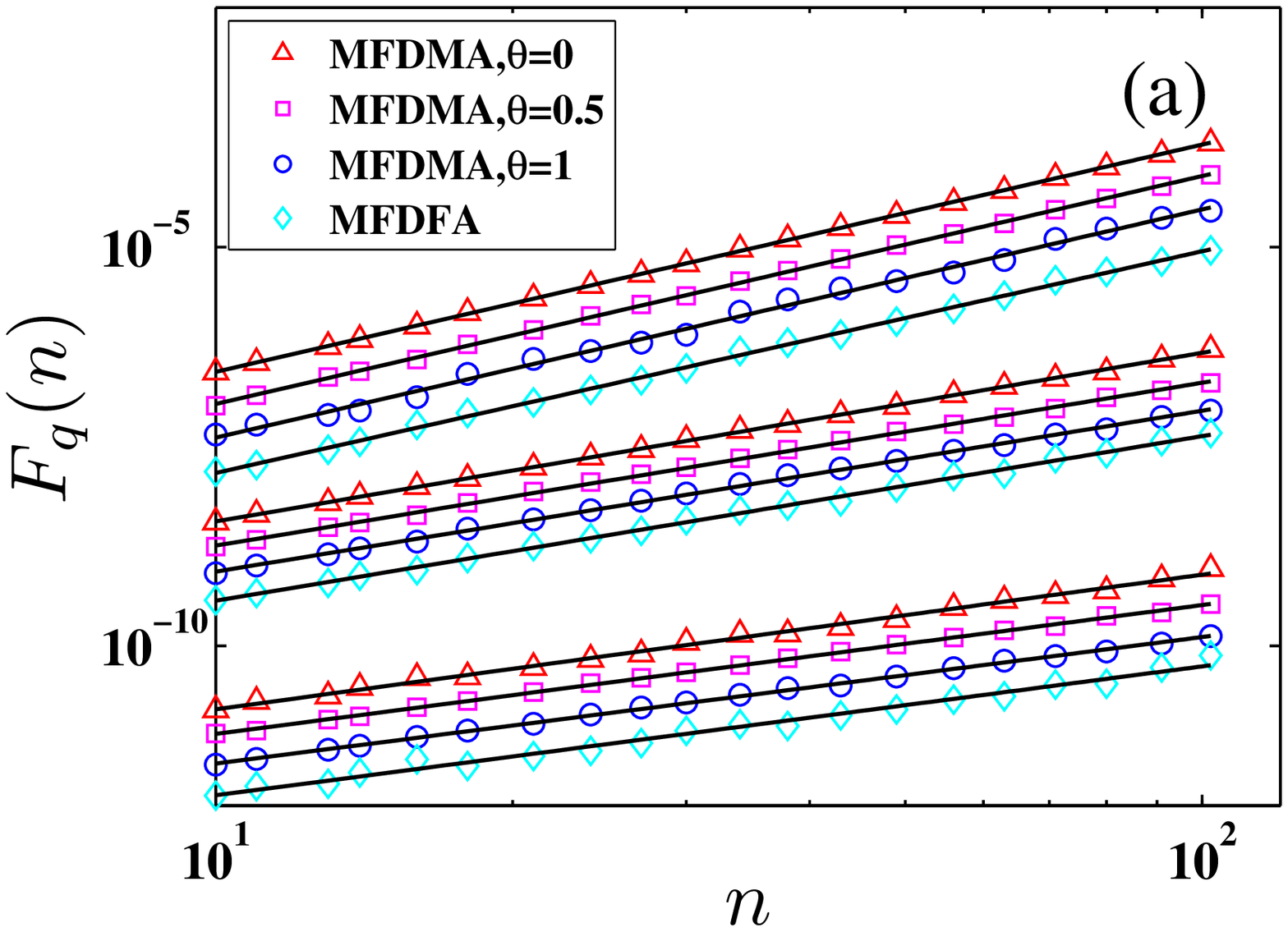}
  \includegraphics[width=8cm]{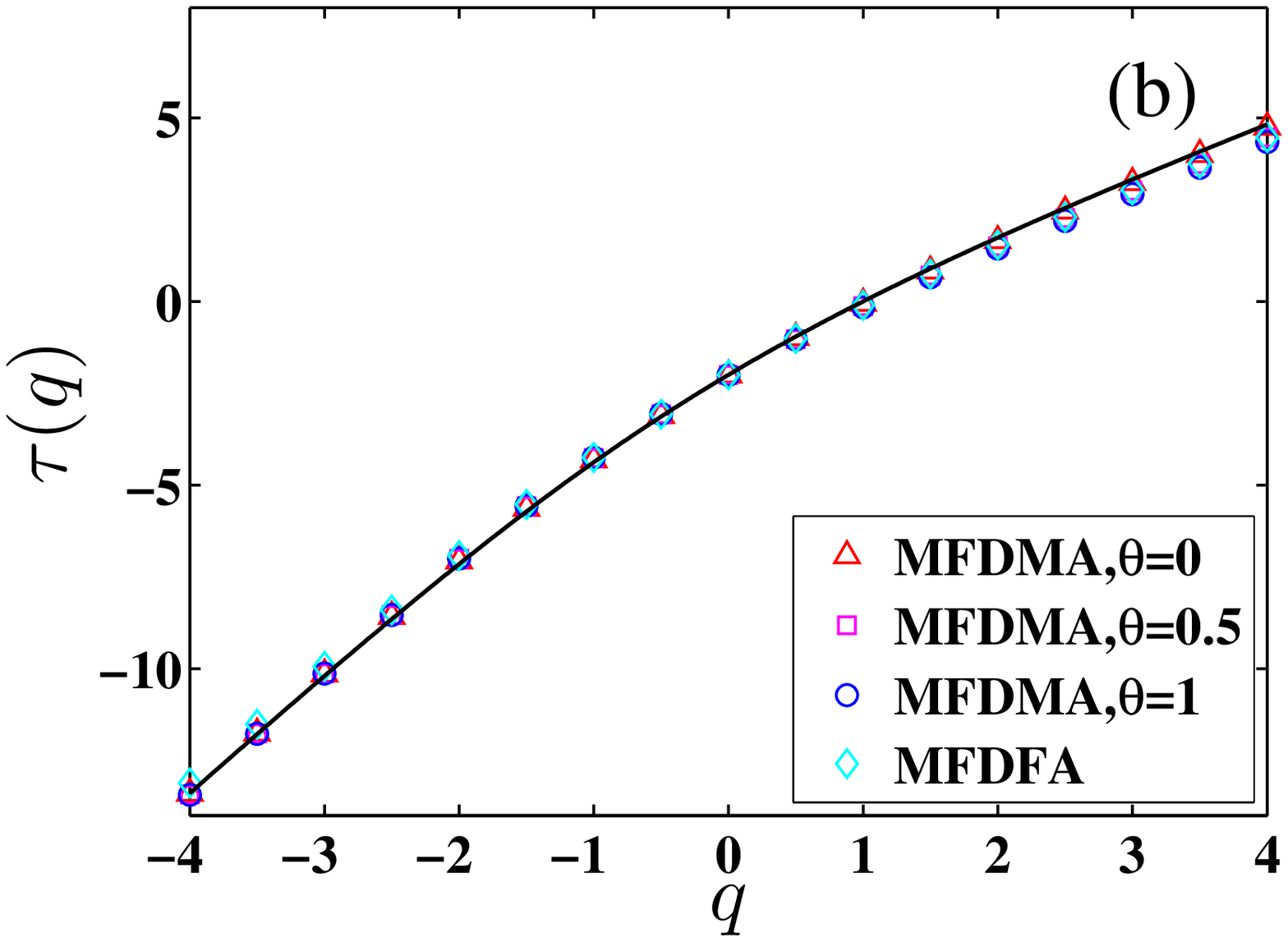}
  \includegraphics[width=8cm]{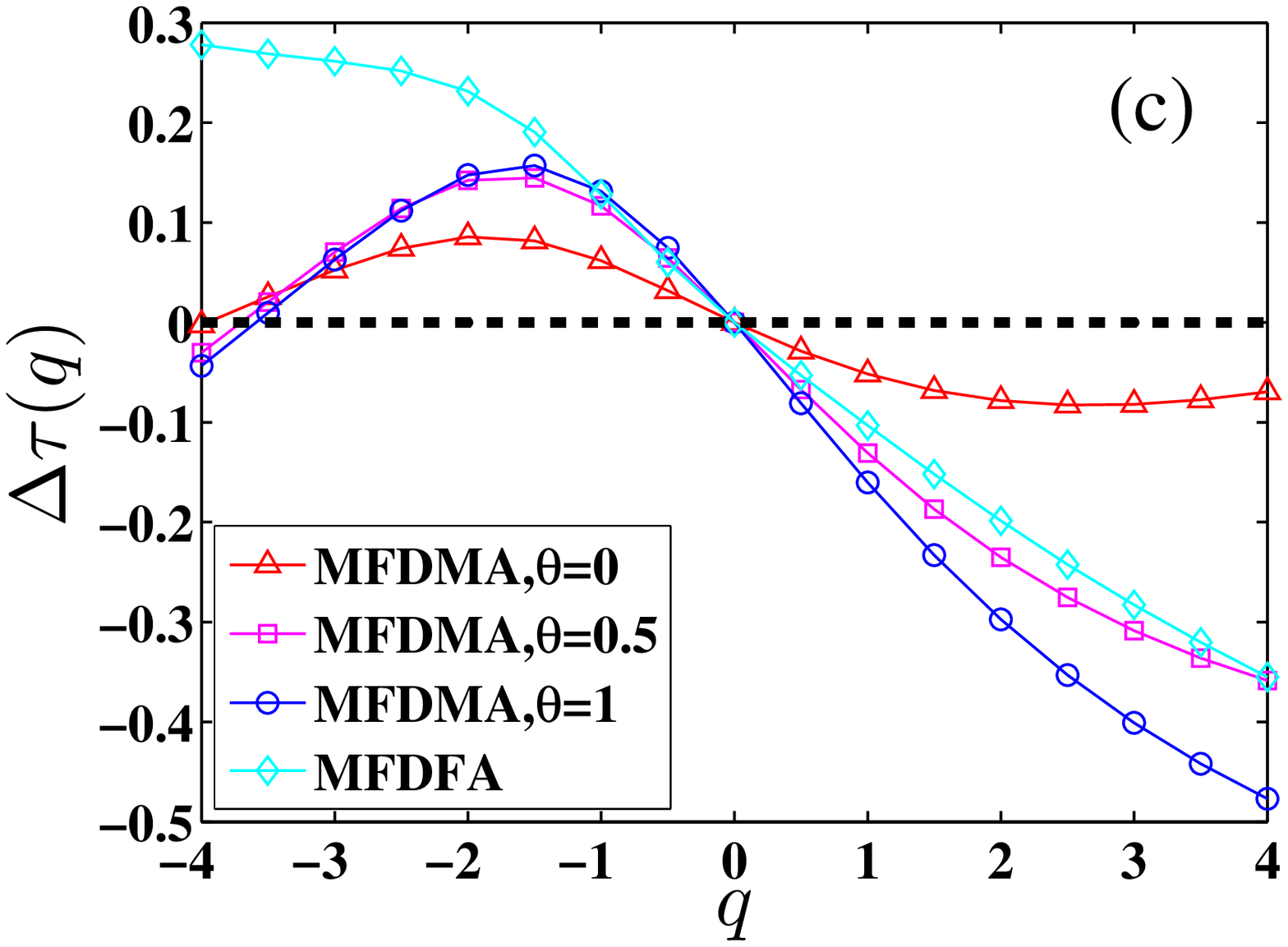}
  \includegraphics[width=8cm]{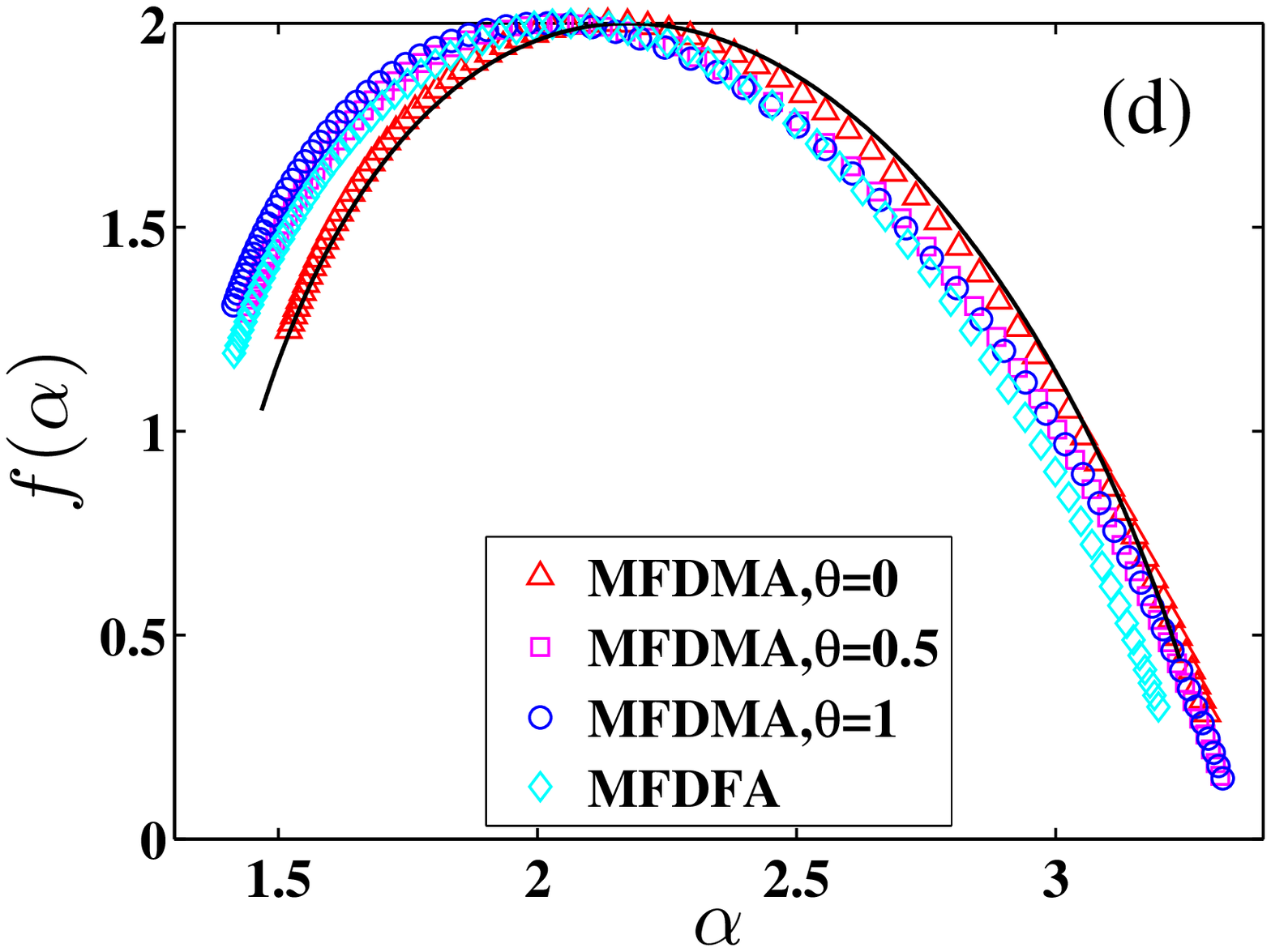}
  \caption{(Color online) Multifractal analysis of the two-dimensional multifractal measure using the three MFDMA algorithms and the MFDFA approach. (a) Power-law dependence of the fluctuation functions $F_q(n)$ with respect to the scale $n$ for $q=-4$, $q=0$, and $q=4$. The straight lines are the best power-law fits to the data. The results have been translated vertically for better visibility. (b) Multifractal mass exponents $\tau(q)$ obtained from the MFDMA and MFDFA methods with the theoretical curve shown as a solid line. (c) Differences $\Delta{\tau}(q)$ between the estimated mass exponents and their theoretical values for the four algorithms. (d) Multifractal spectra $f(\alpha)$ with respect to the singularity strength $\alpha$ for the four methods. The continuous curve is the theoretical multifractal spectrum.}
  \label{Fig:2DMFDMA}
\end{figure*}

In order to investigate the performance of the two-dimensional MFDMA methods, we adopt the multiplicative cascading process to synthesize the two-dimensional multifractal measure. The process begins with a square, and we partition it into four sub-squares with the same size. We then assign four proportions of measure $p_1$, $p_2$, $p_3$ and $p_4$ to them (s.t. $p_1+p_2+p_3+p_4=1$). Each sub-square is further partitioned into four smaller squares and the measure is re-assigned with the same proportions. The procedure is repeated 10 times and we finally generate the two-dimensional multifractal measure with size $1024\times1024$. In the paper, the model parameters are $p_1=0.1$, $p_2=0.2$, $p_3=0.3$, and $p_4=0.4$.

In Fig.~\ref{Fig:2DMFDMA}(a), we depict the fluctuation functions $F_q(n)$ of the two-dimensional multifractal measure for three MFDMA algorithms with $\theta=0$, $\theta=0.5$ and $\theta=1$ described in Sec.~\ref{S2:MFDMA:2D:Algo} and the two-dimensional MFDFA method \cite{Gu-Zhou-2006-PRE}. We find that every $F_q(n)$ function scales excellently as a power law of $n$. Linear least-squares regressions of $\ln[F_n(q)]$ against $\ln(n)$ for each $q$ give the estimates of $h(q)$. Table~\ref{Tb:MFDMA:2D} shows the resultant values of $h(q)$ for $q=-4$, -2, 0, 2, and 4. Similar to the results of the one-dimensional MFDMA and MFDFA methods, we find that the three two-dimensional MFDMA algorithms give better power-law scaling relations than the MFDFA algorithm with smaller standard errors in the parentheses except for the forward MFDMA with $\theta=1$ when $q=-4$. Compared with the theoretical values in the last column of Table \ref{Tb:MFDMA:2D}, the backward MFDMA with $\theta=0$ gives the most accurate estimates and thus has the best performance.

\begin{table}[htp]
 \centering
 \caption{The MFDMA exponents $h(q)$ for $q=-4$, -2, 0, 2, and 4 of the two-dimensional synthetic multifractal measure with the parameters $p_1=0.1$, $p_2=0.2$, $p_3=0.3$, and $p_4=0.4$ using the MFDMA ($\theta=0$, $\theta=0.5$ and $\theta=1$) and MFDFA methods. The numbers in the parentheses are the standard errors of the regression coefficient estimates using the $t$-test at the 5\% significance level.}
 \medskip
 \label{Tb:MFDMA:2D}
 \begin{tabular}{cccccccccccc}
 \hline\hline
  \multirow{3}*[2mm]{$q$}&& \multicolumn{3}{c}{MFDMA}&& \multirow{3}*[2mm]{MFDFA} & \multirow{3}*[2mm]{Analytic} \\  %
  \cline{3-5}
     && $~\theta=0~$ & $\theta=0.5$ & $~\theta=1~$  \\
 \hline
    -4 && 2.850( 7) & 2.857(11) & 2.860(42) && 2.780(20) & 2.849\\
    -2 && 2.534( 6) & 2.505(14) & 2.503(22) && 2.461(23) & 2.577\\
     0 && 2.114(10) & 2.041(13) & 2.017(15) && 2.064(20) & 2.176\\
     2 && 1.829(19) & 1.751(10) & 1.720(10) && 1.769(30) & 1.869\\
     4 && 1.688(25) & 1.615(10) & 1.586(11) && 1.616(43) & 1.705\\
 \hline\hline
 \end{tabular}
\end{table}

In Fig.~\ref{Fig:2DMFDMA}(b), we plot the multifractal scaling exponents $\tau(q)$ estimated from the three MFDMA algorithms and the MFDFA method. The theoretical results are also shown for comparison, which are calculated as follows
\begin{equation}
  \tau_{\rm{th}}(q)=-\frac{\ln(p_1^q+p_2^q+p_3^q+p_4^q)}{\ln2}.
  \label{Eq:2ddma_tau_2}
\end{equation}
According to Fig.~\ref{Fig:2DMFDMA}(b), we find that the four $\tau(q)$ curves estimated from the MFDMA and MFDFA methods are also close to the theoretical values.

To have a better visibility, we plot the difference functions $\Delta\tau(q) = \tau(q)-\tau_{\rm{th}}(q)$ in Fig.~\ref{Fig:2DMFDMA}(b). For positive values of $q$, the algorithms underestimate the $\tau(q)$ values. It is clear that the backward MFDMA with $\theta=0$ performs best, the forward MFDMA with $\theta=1$ performs worst, and the MFDFA outperforms the centered MFDMA with $\theta=0.5$. For negative values of $q$, most $\Delta\tau$ values are larger than 0. We find that the backward MFDMA has the best performance, the MFDFA has the worse performance, and the performance of the centered and forward MFDMA methods are comparable to each other. These findings are further confirmed by the results illustrated in Fig.~\ref{Fig:2DMFDMA}(d) for the multifractal spectra.

\section{Discussion and conclusion}
\label{S1:Summary}

In this paper, we have developed the detrending moving average techniques to make them suitable for the analysis of multifractal measures in one and two dimensions. Extensions to higher dimensions are straightforward. The performance of these MFDMA algorithms is tested based on numerical experiments of synthetic multifractal measures with known theoretical multifractal properties generated according to multiplicative cascading processes. We also present the results of MFDFA for comparison. Our main conclusion is that the backward MFDMA with the parameter $\theta=0$ exhibits the best performance when compared with the centered MFDMA, forward MFDMA, and MFDFA, because it gives better power-law scaling in the fluctuation functions and more accurate estimates of the multifractal scaling exponents and the singularity spectrum.

For the one-dimensional MFDMA version, we find that MFDMA gives a more reliable regression in the log-log plot of the fluctuation function $F_q(n)$ with respect to the scale $n$ than MFDFA. Comparing with the theoretical formulas of $\tau(q)$ and $f(\alpha)$, the backward MFDMA performs best, the centered MFDMA performs worst, and the forward MFDMA outperforms the MFDFA. The backward MFDMA with $\theta=0$ is applied to the analysis of the return time series of SSEC and confirms that the return series exhibits multifractal nature, which is not caused by the fat-tailedness of the return distribution.

For the two-dimensional MFDMA version, we also find that MFDMA gives a more reliable regression than MFDFA when testing on the two-dimensional synthetic multifractal measure. The estimates of $\tau(q)$ and $f(\alpha)$ well agree with the theoretical values for the backward MFDMA with $\theta=0$, which is the best estimator. The centered and forward MFDMA methods with $\theta=0.5$ and $\theta=1$ give a better estimation than MFDFA in the negative range of $q$, while they are worse than MFDFA for positive $q$.

Technically, it is crucial to emphasize that, the window size ($n$ for one-dimensional MFDMA or $n_1\times{n_2}$ for two-dimensional MFDMA) used to determine the moving average function ($\widetilde{y}(i)$ for one-dimensional version or $\widetilde{Y}(i_1,i_2)$ for two-dimensional version) in Step 2 must be identical to the partitioning segment size in Step 4. If the window size is not equal to the segment size, the fluctuation function $F_q(n)$ does not show power-law dependence on the scale $n$, and the multifractal scaling exponent $\tau(q)$ and the multifractal spectrum $f(\alpha)$ estimated from the MFDMA remarkably deviate from the theoretical ones.

Finally, we would like to stress that there are tremendous potential applications of the backward MFDMA method in multifractal analysis, due to the better performance of this algorithm compared with the extensively used MFDFA approach. Possible applications include time series (one-dimensional), fracture surfaces, landscapes, clouds, and many other images possessing self-similar properties (two-dimensional), temperature fields and concentration fields (three-dimensional), and strange attractors in nonlinear dynamics (higher-dimensional). The MATLAB codes are publicly available at http://arxiv.org/abs/1005.0877 for download.

\begin{acknowledgments}
We acknowledge financial supports from the Fundamental Research Funds for the Central Universities and the Program for New Century Excellent Talents in University (NCET-07-0288).
\end{acknowledgments}

\bibliography{E:/Papers/Auxiliary/Bibliography}



\begin{widetext}

\appendix*

\section{MATLAB codes for Multifractal Detrending Moving Average (MFDMA) analysis}

These MATLAB codes are not submitted for publication. They are available as the appendix of the manuscript on the arXiv. Rather than copying directly from the generated PDF file, the readers are advised to download the source files and copy the MATLAB codes from the tex file. Questions and suggestions, if any, should be addressed to gfgu@ecust.edu.cn or wxzhou@ecust.edu.cn.


\lstset{breaklines}
\lstset{language=MATLAB}
\begin{lstlisting}

% GFGU_MFDMA_1D.m

function [n,Fq,tau,alpha,f] = GFGU_MFDMA_1D(x,n_min,n_max,N,theta,q)
%
% The procedure [n,Fq,tau,alpha,f]=GFGU_MFDMA_1D(x,n_min,n_max,N,theta,q) is
% used to calculate the multifractal properties of one-dimensional time series.
%
% Input:
%   x: the time series we considered
%   n_min: the lower bound of the segment size n
%   n_max: the upper bound of the segment size n
%   N: the length of n, that is, the data points in the plot of Fq VS n
%   theta: the position parameter of the moving window
%   q: multifractal order
%
% Output:
%   n: segment size series
%   Fq: q-th order fluctuation function
%   tau: multifractal scaling exponent
%   alpha: multifractal singularity strength function
%   f: multifractal spectrum
%
% The procedure works as follows:
%   1) Construct the cumulative sum y.
%   2) For each n, calculate the moving average function \widetilde{y}.
%   3) Determine the residual e by detrending \widetilde{y} from y.
%   4) Estimate the root-mean-square function F.
%   5) Calculate the q-th order overall fluctuation function Fq.
%   6) Calculate the multifractal scaling exponent tau(q).
%   7) Calculate the singularity strength function alpha(q) and spectrum f(alpha).
%
% Note:
%   1) The window size and the segment size must be identical.
%   2) The lower bound n_min would better be selected around 10.
%   3) The upper bound n_max would better be selected around 10% of the length of
%      time series.
%   4) N would better be seleceted in the range [20,40].
%   5) The parameter theta varies in the range [0,1]. Theta = 0 corresponds to
%      backward MFDMA, and theta = 0.5 corresponds to the centered MFDMA, and
%      theta = 1 corresponds to the forward MFDMA. We recommend theta=0.
%
% Example:
%   [n,Fq,tau,alpha,f]=GFGU_MFDMA_1D(x,10,round(length(x)/10),30,0,-4:0.1:4);
%

if size(x,2) == 1
    x = x';
end

M = length(x);
MIN = log10(n_min);
MAX = log10(n_max);
n = (unique(round(logspace(MIN,MAX,N))))';

% Construct the cumulative sum y
y = cumsum(x);

for i = 1:length(n)
    lgth = n(i,1);

    % Calculate the moving average function \widetilde{y}
    y1 = zeros(1,M-lgth+1);
    for j = 1:M-lgth+1
        y1(j) = mean(y(j:j+lgth-1));
    end

    % Determine the residual e
    e=y(max(1,floor(lgth*(1-theta))):max(1,floor(lgth*(1-theta)))+length(y1)-1)-y1;

    % Estimate the root-mean-square function F
    for k=1:floor(length(e)/lgth)
        F{i}(k)=sqrt(mean(e((k-1)*lgth+1:k*lgth).^2));
    end
end


% Calculate the q-th order overall fluctuation function Fq
for i=1:length(q)
    for j=1:length(F)
        f=F{j};
        if q(i) == 0
            Fq(j,i)=exp(0.5*mean(log(f.^2)));
        else
            Fq(j,i)=(mean(f.^q(i)))^(1/q(i));
        end
    end
end


% Calculate the multifractal scaling exponent tau(q)
for i=1:size(Fq,2)
    fq=Fq(:,i);
    r=regstats(log(fq),log(n),'linear',{'tstat'});
    k=r.tstat.beta(2);
    h(i,1)=k;
end
tau=h.*q'-1;


% Calculate the singularity strength function alpha(q) and spectrum f(alpha)
dx=7;
dx=fix((dx-1)/2);
for i=dx+1:length(tau)-dx
    xx=q(i-dx:i+dx);
    yy=tau(i-dx:i+dx);
    r=regstats(yy,xx,'linear',{'tstat'});
    alpha(i,1)=r.tstat.beta(2);
end
alpha=alpha(dx+1:end);
f=q(dx+1:end-dx)'.*alpha-tau(dx+1:end-dx);


% GFGU_MFDMA_2D.m

function [n,Fq,tau,alpha,f] = GFGU_MFDMA_2D(X,n_min,n_max,N,theta,q)
%
% The procedure [n,Fq,tau,alpha,f]=GFGU_MFDMA_2D(x,n_min,n_max,N,theta,q)
% is used to calculate the multifractal properties of two-dimensional
% multifractal measures.
%
% Input:
%   X: the two-dimensional multifractal measures we considered.
%   n_min: the lower bound of the segment size n
%   n_max: the upper bound of the segment size n
%   N: the length of n, that is, the data points in the plot of Fq VS n
%   theta: the position parameter of the moving window
%   q: multifractal order
%
% Output:
%   n: segment size series
%   Fq: q-th order fluctuation function
%   tau: multifractal scaling exponent
%   alpha: multifractal singularity strength function
%   f: multifractal spectrum
%
% The procedure works as follows:
%   1) For each n, construct the cumulative sum Y in a moving window.
%   2) Calculate the moving average function \widetilde{Y}.
%   3) Determine the residual e by detrending \widetilde{Y} from Y.
%   4) Estimate the root-mean-square function F.
%   5) Calculate the q-th order overall fluctuation function Fq.
%   6) Calculate the multifractal scaling exponent tau(q).
%   7) Calculate the singularity strength function alpha(q) and spectrum f(alpha).
%
% Note:
%   1) The window size and the segment size must be identical.
%   2) The lower bound n_min would better be selected around 10.
%   3) The upper bound n_max would better be selected around 10% of min(size(X)).
%   4) N would better be seleceted in the range [20,40].
%   5) The parameter theta varies in the range [0,1], and we have
%      theta=theta_1=theta_2. Theta = 0 corresponds to backward MFDMA, and
%      theta = 0.5 corresponds to the centered MFDMA, and theta = 1 corresponds to
%      the forward MFDMA. We recommend theta=0.
%   6) In the procedure, we have n=n_1=n_2 for the segment size.
%
% Example:
%   [n,Fq,tau,alpha,f]=GFGU_MFDMA_2D(X,10,round(min(size(X))/10),30,0,-4:0.1:4);
%


N1=size(X,1);
N2=size(X,2);

MIN=log10(n_min);
MAX=log10(n_max);
n=(unique(round(logspace(MIN,MAX,N))))';

for i=1:length(n)
    lgth=n(i,1);

    Y = zeros(N1-lgth+1,N2-lgth+1);
    Y1 = zeros(N1-lgth+1,N2-lgth+1);
    for j = 1:N1-lgth+1
        for k = 1:N2-lgth+1
            Z = X(j:j+lgth-1, k:k+lgth-1);
            Z1 = (cumsum((cumsum(Z))'))';

            % Construct the cumulative sum Y
            Y(j,k)=Z1(end,end);

            % Calculate the moving average function \widetilde{Y}
            Y1(j,k)=mean(Z1(:));
        end
    end

    % Determine the residual e
    x0=1:size(Y,1)-min(floor(lgth*theta),lgth-1);
    y0=1:size(Y,2)-min(floor(lgth*theta),lgth-1);
    x1=size(Y1,1)-length(x0)+1:size(Y1,1);
    y1=size(Y1,2)-length(y0)+1:size(Y1,2);
    e=Y(x0,y0)-Y1(x1,y1);

    % Estimate the root-mean-square function F
    for k1=1:floor(size(e,1)/lgth)
        for k2=1:floor(size(e,2)/lgth)
            E=e((k1-1)*lgth+1:k1*lgth, (k2-1)*lgth+1:k2*lgth);
            F{i}(k1,k2)=sqrt(mean(E(:).^2));
        end
    end
end


% Calculate the q-th order overall fluctuation function Fq
for i=1:length(q)
    for j=1:length(F)
        f=F{j}(:);
        if q(i) == 0
            Fq(j,i)=exp(0.5*mean(log(f.^2)));
        else
            Fq(j,i)=(mean(f.^q(i)))^(1/q(i));
        end
    end
end


% Calculate the multifractal scaling exponent tau(q)
for i=1:size(Fq,2)
    fq=Fq(:,i);
    r=regstats(log(fq),log(n),'linear',{'tstat'});
    k=r.tstat.beta(2);
    h(i,1)=k;
end
tau=h.*q'-2;


% Calculate the singularity strength function alpha(q) and spectrum f(alpha)
dx=7;
dx=fix((dx-1)/2);
for i=dx+1:length(tau)-dx
    xx=q(i-dx:i+dx);
    yy=tau(i-dx:i+dx);
    r=regstats(yy,xx,'linear',{'tstat'});
    alpha(i,1)=r.tstat.beta(2);
end
alpha=alpha(dx+1:end);
f=q(dx+1:end-dx)'.*alpha-tau(dx+1:end-dx);

\end{lstlisting}

\end{widetext}

\end{document}